\documentstyle[12pt,aps,epsf,amsfonts,amssymb]{revtex}
\def\mycap#1{ 
\parbox[h]{\textwidth}{\vskip 0.4cm
\footnotesize \baselineskip 0.mm  #1 } }

\begin{document}
\author{$^{1}${\bf Manuel Houzet, }$^{1}${\bf Alexandre Buzdin,}
and
$^2${\bf %
Miodrag} {\bf L.} {\bf Kuli\'{c}}}

\address{$^{1}$Centre de Physique Mol\'{e}culaire Optique et Hertzienne,\\
Universit\'{e} Bordeaux 1-UMR 5798, CNRS, F-33405 Talence Cedex, France\\
$^{2}$Physikalisches Institut, Theorie III, Universit\"{a}t Bayreuth,\\
D-95440 Bayreuth, Germany}
\title{{\bf Decoupling of superconducting layers in magnetic superconductor RuSr}$%
_{2}${\bf GdCu}$_{2}${\bf O}$_{8}${\bf \ }}
\date{03 March 2001}
\maketitle

\begin{abstract}
We propose the model for magnetic properties of the magnetic superconductor
RuSr$_{2}$GdCu$_{2}$O$_{8}$, which incorporates the theory of the
superconducting/ferromagnetic multilayers. The transition line $T_{d}(h)$,
on which the Josephson coupled superconducting planes are decoupled, i.e. $%
j_{c}(T_{d})=0$, is calculated as a function of the exchange energy $h$. As
the result of this decoupling a nonmonotonic behavior of magnetic
properties, like the lower critical field $H_{c1}$, Josephson plasma
frequency, etc. is realized near (or by crossing) the $T_{d}(h)$ line. The
obtained results are used in analyzing the newly discovered
antiferromagnetic ruthenocuprate RuSr$_{2}$GdCu$_{2}$O$_{8}$ with possible
weak ferromagnetic order in the RuO planes.
\end{abstract}

\newpage

\section{\ Introduction}

The physics of magnetic superconductors is interesting due to competition of
magnetic order and singlet superconductivity in bulk materials. The problem
of their coexistence was first set up theoretically in the pioneering work
by V. L. Ginzburg \cite{Ginzburg} in 1956, while the experimental progress
in the field begun after the discovery of ternary rare earth (RE) compounds
(RE)Rh$_{4}$B$_{4}$ and (RE)Mo$_{6}$X$_{8}$ (X=S,Se) \cite{Maple} with a
regular distribution of localized RE magnetic moments. It turned out that in
many of these systems superconductivity (with the critical temperature $%
T_{c} $) coexists rather easily with antiferromagnetic (AF) order (with the
critical N\'{e}el temperature $T_{N}$), where usually the situation with $%
T_{N}<T_{c}$ is realized \cite{Maple}. Due to their antagonistic characters
singlet superconductivity and ferromagnetic order cannot coexist in bulk
samples with realistic physical parameters. However, under certain
conditions the ferromagnetic order is transformed, in the presence of
superconductivity, into a spiral or domain-like structure - depending on the
type and strength of magnetic anisotropy in the system \cite
{BuBuKuPaAdvances}. As the result of this competition, these two orderings
coexist in a limited temperature interval $T_{c2}<T<T_{m}$ (the reentrant
behavior) in ErRh$_{4}$B$_{4}$ and HoMo$_{6}$S$_{8}$, or even down to $T=0%
\mathop{\rm K}%
$ in HoMo$_{6}$Se$_{8},$ where $T_{m}$\ is the critical temperature for the
existence of the inhomogeneous magnetic order. The coexistence region in ErRh%
$_{4}$B$_{4}$ is narrow where $T_{c}=8.7%
\mathop{\rm K}%
$, $T_{m}\approx 0.8%
\mathop{\rm K}%
$, $T_{c2}\approx 0.7%
\mathop{\rm K}%
$, while for HoMo$_{6}$S$_{8}$ it is even narrower with $T_{c}=1.8%
\mathop{\rm K}%
$, $T_{m}\approx 0.74%
\mathop{\rm K}%
$, $T_{c2}\approx 0.7%
\mathop{\rm K}%
$ - see Refs. \cite{Maple}, \cite{BuBuKuPaAdvances}. In most of the new
quaternary rare-earth compounds (RE)Ni$_{2}$B$_{2}$C the antiferromagnetic
order and superconductivity coexist up to $T=0%
\mathop{\rm K}%
$ \cite{Chang}, while in HoNi$_{2}$B$_{2}$C an additional oscillatory
magnetic structure is realized in a limited temperature interval. This
oscillatory magnetic structure competes strongly with superconductivity
giving rise\ to reentrant behavior in this compound \cite{DetBraun}.
Recently the Pobell%
\'{}%
s group in Bayreuth \cite{Pobell} made a remarkable discovery of the
coexistence of superconductivity and nuclear magnetic order in AuIn$_{2}$
with $T_{c}=0,207%
\mathop{\rm K}%
$ and $T_{m}=35\mu 
\mathop{\rm K}%
$. This exciting phenomenon was explained in Ref. \cite{KuBuBu} where it is
argued that superconductivity can coexist either with spiral or domain-like
nuclear magnetic ordering only, depending on the strength of magnetic
anisotropy in this cubic system. Important contribution to the physics of
magnetic superconductors has been done in Ref. \cite{Buzdin}, where for the
first time was proposed the coexistence of weak-ferromagnetism and
superconductivity. In such a case the spontaneous vortex state due to
weak-ferromagnetism is also possible.

We point out that in the above cited magnetic superconductors the exchange
interaction (between localized magnetic moments and conduction electrons)
influences superconductivity much stronger than the electromagnetic
interaction. The latter is due to the localized magnetic moments which
create dipolar magnetic field, thus affecting the orbital motion of
superconducting electrons.

Recently, a new class of magnetic superconductors based on layered
perovskite ruthenocuprate compound RuSr$_{2}$GdCu$_{2}$O$_{8}$ comprising CuO%
$_{2}$ bilayers and RuO monolayers has been synthesized \cite{Braun}. This
compound belongs also to the class of high-$T_{c}$ superconductors (HTS). A
subsequent study of transport and magnetic properties has revealed that it
exhibits some kind of ferromagnetic order at the critical temperature $%
T_{N}=(133-137)%
\mathop{\rm K}%
$. The polarized neutron scattering measurements \cite{Lynn} show that the
magnetic structure (which appears at $T_{N}$) is predominantly
antiferromagnetic with a Ru magnetic moment $\mu _{\text{Ru}}\approx 1.18\mu
_{B}$ along the c-axis at low temperature. The same measurements put an
upper limit $\sim 0.1\mu _{B}$ to any net ferromagnetic zero-field Ru
moment. Concerning the last point the important results came from
magnetization measurements first reported in Ref. \cite{Braun}, which show a
hysteresis loop and remanent magnetization. The latter hints to existence of
a ferromagnetic component in the system. Recent magnetization measurements
on RuSr$_{2}$EuCu$_{2}$O$_{8}$ \cite{Williams} give evidence for a small
ferromagnetic component, which lies probably parallel to the RuO plane, with
the magnetic moment (per Ru) $\sim 0.05\mu _{B}$ at $5%
\mathop{\rm K}%
$ consistent with the neutron scattering data \cite{Lynn}. Note that the
smaller value of magnetic moment ($0.05\mu _{B}$) in this compound tells us
that in the Gd-compound some admixture of the large Gd moment might take
place. This conclusion is also confirmed by the zero-field muon spin
rotation (ZF-$\mu $SR) measurements \cite{Bernhard} which provide important
evidence that the magnetic order is homogeneous on a microscopic scale and
accounts for most of the sample volume. At lower temperatures the
superconductivity sets in at $T_{c}=(35-45)%
\mathop{\rm K}%
$ without affecting the AF order \cite{Lynn}, \cite{Bernhard} notably. This
fact means that superconductivity - which is realized predominantly in the
CuO$_{2}$ planes, and magnetic order - which is present only in the RuO
planes, interact rather weakly, i.e. these two orders are separated
spatially. Recently, it was reported \cite{Klamut} that in Ru$_{1-x}$Sr$_{2}$%
GdCu$_{2+x}$O$_{8-y}$ the highest superconducting critical temperature
reaches $72%
\mathop{\rm K}%
$ for $x=0.3-0.4$, while there is no sign of the weak-ferromagnetic (WF)
component in the RuO planes.

It seems that RuSr$_{2}$GdCu$_{2}$O$_{8}$ has very interesting magnetic
properties, which might result in the absence of Meissner phase in some
samples \cite{Chu}, \cite{Felner}, while in some others it is realized \cite
{Braun}, \cite{Bernhard2}. (This problem will be briefly discussed in
Section IV.)

In this paper we propose a model of layered magnetic superconductor with
weak-ferromagnetism, which might be relevant for the RuSr$_{2}$GdCu$_{2}$O$%
_{8}$ compound - the SWF model. This model, studied in Section II, assumes
the existence of S/F multilayers with small hopping parameter $t$ between S
(superconducting) and F (ferromagnetic) planes along the $c$-axis , i.e. $%
t<T_{c}$. As a result the small $t$ gives rise to an effective Josephson
coupling current $j_{c}$ between superconducting planes. It turns out that $%
j_{c}$ is suppressed by the exchange field - present in the F-plane only,
which causes drastic changes of magnetic properties. The Gibbs free-energy $%
{\cal G}$ of such a magnetic superconductor with both AF and WF orderings in
external magnetic field ${\bf H}$ is formulated in Section III. Based on it
the lower critical field $H_{c1}$ is also studied there. The estimation of
theoretical parameters of the SWF model from the experimental results in RuSr%
$_{2}$GdCu$_{2}$O$_{8}$ is done in Section IV, where the obtained results
are discussed too.

\section{Model for S/F atomic multilayer and Josephson current}

As was mentioned above we consider the magnetic superconductor RuSr$_{2}$GdCu%
$_{2}$O$_{8}$ as a prototype for S/F atomic multilayers by assuming good
conduction in CuO$_{2}$ planes - with the quasiparticle spectrum $\xi _{S}(%
{\bf p})$, and a small hopping parameter $t<T_{c}$ between the S- and
F-planes (i.e., along the $c$-axis). The second assumption is related to the
existence of WF order (with the magnetization ${\bf M}$ lying in the RuO
planes) which gives rise to an effective exchange field parameter ${\bf h}=h%
{\bf e}_{ab}$. The latter affects spins of conduction electrons with the
dispersion $\xi _{F}({\bf p})$ moving in the normal conducting RuO planes.
(The $ab$-plane is sometimes labeled by the $xy$-plane.) The parameter $h$
can be related to an effective spontaneous spin $S_{%
\mathop{\rm eff}%
}$ (magnetization normalized to saturation magnetization) in the $ab$-plane,
i.e. $h=J^{ab}S_{%
\mathop{\rm eff}%
}$ - see also Section IV.

The electronic part of the SWF model is similar to the model in the Ref. 
\cite{Andreev} and in what follows the same notation is used. According to
this model the elementary cell of the superlattice consists of one
superconducting and one ferromagnetic layer which are both metallic. For
simplicity it is supposed here that both layers have similar quasiparticle
energy spectra, i.e. $\xi ({\bf p})\equiv (\xi _{S}({\bf p})\approx \xi _{F}(%
{\bf p}))$. It is also assumed, that the superconductivity is realized in
S-planes (CuO$_{2}$ planes) with pairing coupling $g({\bf p})$ (having in
mind application to the HTS compound RuSr$_{2}$GdCu$_{2}$O$_{8}$ the clean
limit, $\xi _{0}\ll l$, is supposed). The Hamiltonian of the system is given
by

\begin{eqnarray}
H &=&H_{0}+H_{int1}+H_{int2},\nonumber \\
H_{0} &=&\sum_{{\bf p},n,i,\sigma }\xi ({\bf p})a_{n,i,\sigma }^{\dagger }({\bf %
p)}a_{n,i,\sigma }^{{}}({\bf p)}+t\left[ a_{n,1,\sigma }^{\dagger }({\bf p)}%
a_{n,-1,\sigma }^{{}}({\bf p)}+
a_{n+1,-1,\sigma }^{\dagger }({\bf p)}%
a_{n,1,\sigma }^{{}}({\bf p)}+h.c.\right] \nonumber \\ 
H_{int1} &=&{\frac{1}{2}}\sum_{{\bf p}_{1},{\bf p}_{2},n,\sigma }g({\bf p}%
_{1}-{\bf p}_{2})a_{n,1,\sigma }^{\dagger }({\bf p}_{1})a_{n,1,-\sigma
}^{\dagger }(-{\bf p}_{1})a_{n,1,-\sigma }^{{}}(-{\bf p}_{2})a_{n,1,\sigma
}^{{}}({\bf p}_{2}),  \nonumber \\
H_{int2} &=&-\sum_{{\bf p},n,\sigma }h\sigma a_{n,-1,\sigma }^{\dagger }(%
{\bf p)}a_{n,-1,\sigma }^{{}}({\bf p)},  \nonumber
\end{eqnarray}
where $a_{n,i,\sigma }^{\dagger }({\bf p)}$ is the creation operator of an
electron with spin $\sigma $ (the quantization axis is parallel to the $ab$%
-plane) in the $n$-th elementary cell and momentum ${\bf p}$ in the layer $i$
is parallel to the $ab$-plane, where $i=1$ for the S layer, and $i=-1$ for
the F layer. Since the obtained results below are qualitatively similar for
s- and d-wave pairing, the calculations were done for s-wave pairing where $%
g({\bf p})=g_{0}$ is constant, while quantitative changes due to d-wave
pairing are discussed below and in Section IV.

By assuming that the order parameter changes from cell to cell in the manner 
$\Delta _{n}=|\Delta |e^{i\varphi _{n}}$ (with $\varphi _{n}=kn$ in absence
of orbital effects) the quasiparticle Green's functions are obtained in the
standard way \cite{Andreev}. The self-consistency equation for the order
parameter $|\Delta |$ reads \cite{Andreev}

\begin{equation}
\frac{1}{\Lambda }=T\sum_{\omega }\int_{-\infty }^{\infty }d\xi
\int_{0}^{2\pi }\frac{dq}{2\pi }{\frac{\widetilde{\omega }_{+}\widetilde{%
\omega }_{-}}{|\Delta |^{2}\widetilde{\omega }_{+}\widetilde{\omega }%
_{-}-(\omega _{-}\widetilde{\omega }_{-}-|{\cal T}_{q+k}|^{2})(\omega _{+}%
\widetilde{\omega }_{+}-|{\cal T}_{q}|^{2})},}  \label{celf}
\end{equation}
where $\Lambda =g_{0}\rho (0)$ and $\rho (0)=m_{\parallel }/2\pi $ is the
electron density of states at the Fermi level in the normal state, $\omega
_{\pm }=i\omega \pm \xi (p)$, $\widetilde{\omega }_{\pm }=\omega _{\pm }+h$, 
$\omega =\pi T(2n+1)$. The quasimomentum $q$ lies in the direction
perpendicular to the layers, and ${\cal T}_{q}=2t\cos (q/2)e^{iq/2}$ and $%
{\cal T}_{q+k}=2t\cos ((q+k)/2)e^{i(q+k)/2}$.

The free energy ${\cal F}$ in the superconducting state is obtained by using
the following relation 
\begin{equation}
\frac{\partial {\cal F}}{\partial |\triangle |}=\frac{|\triangle |}{\Lambda }%
-{\frac{T\rho (0)}{2\pi }}\sum_{\omega }\int_{0}^{\infty }\int_{0}^{2\pi
}d\xi dqF_{11}^{\dagger },  \label{free}
\end{equation}
where the expression for $F_{11}^{\dagger }$ is obtained in \cite{Andreev}.

In order to study transport and magnetic properties in magnetic field we
need to know the supercurrent $j_{z}$ flowing across the layers (along the $%
c $-axis in RuSr$_{2}$GdCu$_{2}$O$_{8}$). In this case, the vector potential 
${\bf A}_{z}=A_{z}{\bf e}_{z}$ enters the Hamiltonian through the
substitution $t\rightarrow te^{\pm iedA_{z}/c},$ where $d$ is S-F interlayer
distance, and the part of the Hamiltonian depending on $A_{z}$ is given by 
\begin{equation}
H_{A}=\sum_{{\bf p},n,i,\sigma }t[a_{n,1,\sigma }^{\dagger }({\bf p)}%
a_{n,-1,\sigma }^{{}}({\bf p)}e^{iedA_{z}/c}+a_{n+1,-1,\sigma }^{\dagger }(%
{\bf p)}a_{n,1,\sigma }^{{}}({\bf p)}e^{iedA_{z}/c}+h.c.].
\end{equation}
The supercurrent across the planes is obtained by the standard procedure 
\begin{equation}
j_{z}=-\frac{c}{2d}{\frac{\delta H_{A}}{\delta A_{z}}.}  \label{jz}
\end{equation}
Note that the Josephson supercurrent in the S/F superlattice is carried by
Andreev bound states, similarly to the S/N case. In S layer the supercurrent
is carried by Cooper pairs, but in N layer it flows via quasiparticles,
which recondense in the next S layer; bound states represent this process 
\cite{Furusaki}.

In the case of small hopping parameter $t\ll T_{c}$ the Josephson current
along the $c$-axis is obtained in leading order (proportional to $t^{4}$) by
standard perturbation theory. After the integration over the energy $\xi $
it reads

\begin{eqnarray}
j_{z} &=&4e\pi |\Delta |^{2}t^{4}\rho (0)T\sum_{\omega >0}\{2\omega \frac{%
5h^{4}+6h^{2}|\Delta |^{2}+|\Delta |^{4}-4h^{2}\omega ^{2}}{R^{2}(\omega
)(\omega ^{2}+h^{2})}  \label{josep} \\
&&{-}\frac{|\Delta |^{2}+h^{2}}{R(\omega )\Omega ^{3}(\omega )}-2\frac{%
(|\Delta |^{2}+h^{2})^{2}-4h^{2}\omega ^{2}}{R^{2}(\omega )\Omega (\omega )}%
\}\sin \left( {k}\right)  \nonumber \\
&=&j_{c}\sin {k,}  \nonumber
\end{eqnarray}
where $R(\omega )=(|\Delta |^{2}+h^{2})^{2}+4h^{2}\omega ^{2}$ and $\Omega
(\omega )=\sqrt{\omega ^{2}+|\Delta |^{2}}$. As in the standard Josephson
effect, the supercurrent $j_{z}$ is proportional to $\sin {k}$, $k$ being
the phase difference between $n$- and $n+1$ S- layers.

In what follows we calculate numerically the critical current $j_{c}$ (in
Eq.(\ref{josep})) at any point of the phase diagram $\left( T,h\right) $ by
replacing $|\Delta |\rightarrow \Delta _{0}(T)$, where $\Delta _{0}(T)$ is
given by the BCS theory. The latter is correct due to the smallness of $t,$
in which case $T_{c}$ is practically unaffected by the exchange field, i.e. $%
T_{c}\approx T_{c0}$ up to the second order terms in $t/T_{c0}$.

From Eq.(\ref{josep}) it comes out in particular, that near $T_{c}$ and for $%
h=0$ one has $j_{c}>0$, while $j_{c}<0$ for $h\gg T_{c}$. The change of sign
of $j_{c}$ (near $T_{c}$), which corresponds to the transition from $k=0$ to 
$k=\pi $ in the ground state, occurs at $h_{c}=3.77T_{c0}$, in accordance
with the calculation in \cite{Andreev}. At low temperatures, $T\rightarrow 0$%
, $j_{c}$ goes to zero at $h/\Delta _{0}(0)\simeq 1/2$, which just
corresponds to $h_{c0}=0.87T_{c0}$ at $T=0,$ again in accordance with Ref. 
\cite{Andreev}. Note that the same approach if applied to d-wave pairing 
\cite{Prokic} gives $h_{c0}^{(d)}=0.6T_{c0}$ at $T=0$. The sign-change of $%
j_{c}$ is related to the transition from the ``$0$''-phase to ``$\pi $%
''-phase. This transition goes smoothly if we take into account the higher
order term ($\sim t^{8}\cos 2k$) in the free-energy, in fact it means that
the width of the region $\Delta h$ where the transition from ``$0$''-phase
to ``$\pi $''-phase occurs is of the order of $\Delta h\sim t^{4}/T_{c0}^{3}$%
. In the case of weak hopping $t\ll T_{c0}$ this region is very narrow and
we may define the decoupling line $j_{c}(T_{d},h)=0$ which results in the $%
(T_{d},h)$ phase diagram shown in Fig.1.
\newline
\epsfysize=3.0in 
\hspace*{0.5cm} 
\vspace*{0cm} 
\epsffile{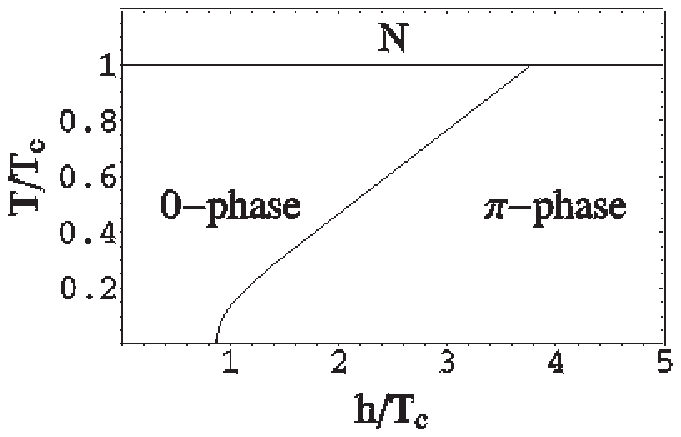} 
\newline
\mycap{{\bf FIGURE~1.}The $(T,h)$-phase diagram for the case $t\ll T_{c0}$.
The Josepson decoupling line $j_{c}(T,h)=0$ - black line }

The temperature dependence of the Josephson penetration depth $\lambda _{J}=%
\sqrt{c\phi _{0}/8\pi ^{2}\left| j_{c}\right| (2d)}$ \cite{Lev}$,$ where $%
\phi _{0}$ is the flux quantum, is shown in Fig.2 for various $h\neq 0$.
Here $2d$ is the period of the multilayer. One should note its nonmonotonic
behavior if $h\neq 0$, particularly when $h\sim T_{c}$. Based on these
results one can analyze some magnetic properties, like the lower critical
field $H_{c1}$ in the $ab$-plane.
\newline
\epsfysize=3.0in 
\hspace*{0.5cm} 
\vspace*{0cm} 
\epsffile{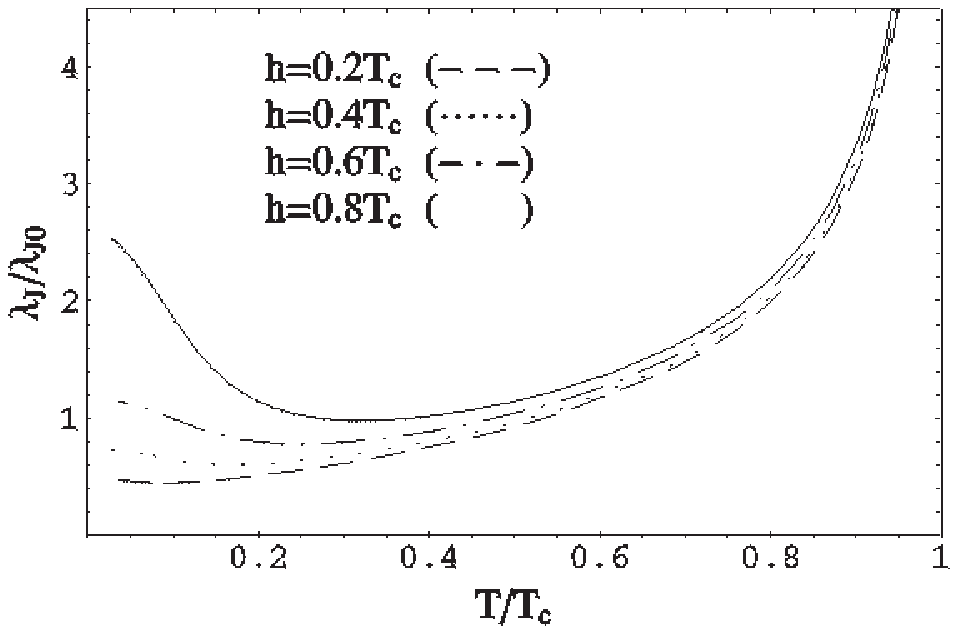} 
\newline
\mycap{{\bf FIGURE~2.} The $T$-dependence of the Josephson penetration 
depth $\lambda _{J}(T,h)$ for various $h.$ We defined 
$\Delta _{0}=1.76T_{c0}$, $j_{c0}=e\rho
(0)t^{4}/\Delta _{0}^{2},$ and $\lambda _{J0}=\sqrt{c\phi _{0}/16\pi
^{2}dj_{c0}}.$ }

\section{Gibbs energy and in-plane critical field $H_{c1}$}

\subsection{Gibbs energy}

In order to calculate the lower critical field $H_{c1}$ (and the possible
absence of Meissner phase \cite{Chu}, \cite{Felner}) we need the Gibbs
energy functional ${\cal G}$. Having in mind the application to the RuSr$%
_{2} $GdCu$_{2}$O$_{8}$ we assume, according to the neutron scattering data 
\cite{Lynn}, that in the magnetic subsystem (F-layers coinciding with RuO
planes) AF order with spins along the $c$-axis is realized at $T_{N}\gg
T_{c} $. The AF order parameter is ${\bf L}=L_{z}{\bf e}_{z}$. The
magnetization measurements \cite{Braun}, \cite{Williams} imply WF order with
the magnetization lying (most probably) parallel to the $ab$-planes and with
the effective moment, $\left| {\bf M}\right| /n_{\text{Ru}}=\mu _{%
\mathop{\rm eff}%
}<0.1\mu _{B}$, where $n_{\text{Ru}}$ is the density of the magnetic Ru
ions. In order to construct the magnetic free-energy it is necessary to know
the symmetry of the system as well as orientations of the easy axis in
different sublattices. So, for instance if the magnetic anisotropy energy on
different sublattices are unequal, then one expects the WF order to be
realized. However, at present there are no sufficient experimental data on
the local lattice distortion which might favor WF order and accordingly the
preferred direction of ${\bf M}$. The above discussed experiments \cite{Lynn}%
, \cite{Bernhard}, \cite{Braun}, \cite{Williams} suggest only that ${\bf M}$
is in the $ab$-plane, i.e. ${\bf M}=M_{\eta }{\bf e}_{\eta }$, where ${\bf e}%
_{\eta }$ is in the $ab$-plane. In the following we define the $x$-axis by $%
{\bf e}_{x}\equiv {\bf e}_{\eta }$. As the result of this analysis the SWF
model contains the following order parameters: $L_{z}$ for the AF order, $%
M_{x}$ for the WF order and $\Delta _{n}(x,y)$ for the S order.

In the applied magnetic field ${\bf H}$ the Gibbs energy of the layered
magnetic superconductor reads - see also \cite{BuBuKuPaAdvances}, \cite
{Buzdin}, 
\begin{eqnarray}
{\cal G}\left[ \Delta _{n},{\bf L,M},{\bf B};{\bf H}\right] &=&\int
dV\left( {\cal F}_{M}\left[ {\bf L},{\bf M}\right] +\frac{({\bf B}-4\pi {\bf %
M)}^{2}}{8\pi }-\frac{{\bf BH}}{4\pi }\right)  \label{Gibbs}  \\ 
&{+}&\sum_{n}\left( 2d\right) \int
dxdy{\cal F}_{S}\left[ \Delta _{n},{\bf A}\right] \nonumber , 
\end{eqnarray}
where $\Delta _{n}\equiv \Delta _{n}(x,y)$, and ${\bf L,}$ ${\bf M},$ ${\bf B%
}$ are also coordinate dependent. The magnetic field ${\bf B}=%
\mathop{\rm rot}%
{\bf A}$ is due to the dipolar field created by the magnetic moments, the
external magnetic field and the superconducting screening current. The
vector potential ${\bf A}={\bf A}_{ab}{\bf +A}_{c}=A_{ab}{\bf e}_{ab}+A_{z}%
{\bf e}_{z}$ contains the component, $A_{ab}$, in the $ab$-plane, and $A_{z}$
along the $c$-axis.

The magnetic free-energy density functional ${\cal F}_{M}\left[ {\bf L},{\bf %
M}\right] $ which mimics the experimental results in RuSr$_{2}$GdCu$_{2}$O$%
_{8}$ with ${\bf L}=L_{z}{\bf e}_{z}$ and ${\bf M}=M_{x}{\bf e}_{x}$ is
given by the following phenomenological expression 
\begin{equation}
{\cal F}_{M}\left[ {\bf L},{\bf M}\right] =\frac{\alpha }{2}{\bf L}^{2}+%
\frac{\beta }{4}{\bf L}^{4}+\frac{\delta }{2}{\bf M}^{2}-\gamma
L_{z}M_{x}+F_{a}({\bf L},{\bf M})+F_{g}(\nabla {\bf L},\nabla {\bf M)+...}
\label{mfree}
\end{equation}
Although this expression is quantitatively correct near the AF transition at 
$T_{N}$ it is also suitable for semiquantitative analysis even below
superconducting transition temperature $T_{c}$, due to the smallness of $%
{\bf M}$ and $\gamma $. The first two terms describe the AF order ($\alpha
=\alpha ^{\prime }(T-T_{N})<0$, $\beta >0$), the third ($\delta >0$) and
fourth ($\sim \gamma $) terms describe the induced WF by the AF order. The
parameters $\alpha $ and $\delta $ are due to the exchange interaction
(between Ru spins in RuSr$_{2}$GdCu$_{2}$O$_{8}$), where one has $\alpha
^{\prime }\sim 1/\theta _{em}$ and $\delta \sim T_{N}/\theta _{em}$ with $%
\theta _{em}=2\pi \mu _{B}^{2}\sim 1%
\mathop{\rm K}%
$. The unknown anisotropy term ${\cal F}_{a}$ fixes the direction of ${\bf L}
$ and ${\bf M}$, i.e. $L_{z}$, $M_{x}$.

Since in the following we analyze the lower critical field along the $ab$%
-plane, with characteristic length-scales $\lambda _{ab}$,$\lambda _{J}\gg
\xi _{ab},d_{\text{Ru-Ru}}$, where $\xi _{ab}$ is the coherence length and $%
d_{\text{Ru-Ru}}$ is the Ru-Ru distance, it is justified to omit the
gradient term ${\cal F}_{g}(\nabla {\bf L},\nabla {\bf M)}$. By minimizing $%
{\cal F}_{M}\left[ {\bf L},{\bf M}\right] $ with respect to $L_{z}$ and $%
M_{x}$ one gets (at temperatures $T_{c}<T<T_{N}$) 
\begin{equation}
M_{x}^{0}=\frac{\gamma }{\delta }L_{z}.  \label{wf}
\end{equation}
The neutron scattering and magnetization measurements give limits for $%
(L_{z}/n_{\text{Ru}})\sim (1-1.2)\mu _{B}$, and $(M_{x}^{0}/n_{\text{Ru}%
})<(0.05-0.1)\mu _{B}$, which implies an upper limit for $\gamma $, i.e. $%
(\gamma /\delta )\lesssim (0.05-0.1)$.

According to experiments \cite{Lynn}, \cite{Bernhard} the AF (and WF)
ordering is practically unaffected by the appearance of superconductivity,
then it is reasonable to neglect the effect of superconductivity 
on the exchange (RKKY)
interaction in ${\cal F}_{S}\left[ \Delta _{n},{\bf A}\right] $. Therefore
we keep in ${\cal F}_{S}\left[ \Delta _{n},{\bf A}\right] $ the
electromagnetic interaction between superconducting electrons and magnetic
order only 
\begin{equation}
{\cal F}_{S}\left[ \Delta _{n},{\bf A}\right] ={\cal F}_{S}\left[ \left|
\Delta _{n}\right| ,0\right] +\left( \frac{4\pi \lambda _{ab}}{c}\right) ^{2}%
{\bf j}_{ab}^{2}+\frac{j_{c}\Phi _{0}}{2\pi c}(1-\cos \chi _{n,n+1}),
\label{sfree}
\end{equation}
where $F_{S}\left[ \left| \Delta _{n}\right| ,0\right] $ is the condensation
energy and $\Delta _{n}=\left| \Delta _{n}\right| \exp (i\varphi _{n})$. Note 
that the exchange interaction between conduction electrons and localized Ru
moments affects superconductivity by renormalizing $j_{c}$, which is function 
of h.

The current in the $ab$-plane ${\bf j}_{ab}$ reads 
\begin{equation}
{\bf j}_{ab}=-\frac{c}{4\pi \lambda _{ab}^{2}}({\bf A}_{ab}-\frac{\phi _{0}}{%
2\pi }\nabla _{ab}\varphi _{n}),  \label{jab}
\end{equation}
where $\lambda _{ab}$ is the bulk London penetration depth in the $ab$
superconducting layers (we assumed $\lambda _{a}=\lambda _{b}\equiv \lambda
_{ab}$). The last term depends on the gauge invariant phase $\chi _{n,n+1}$ 
\begin{equation}
\chi _{n,n+1}=\varphi _{n+1}-\varphi _{n}-\frac{2\pi A_{z}d}{\phi _{0}},
\label{gphase}
\end{equation}
which characterizes the effective Josephson coupling between two neighboring
S-planes with the distance $2d$. It is due to the hopping between S- and
F-planes and $j_{c}\equiv $ $j_{c}(T,h)$ is determined by Eq.(\ref{josep}).

\subsection{Lower critical field $H_{c1}$}

Let us calculate the lower critical field $H_{c1}^{ab}$ for the case when
the magnetic field ${\bf H}$ and the single vortex are along the
magnetization ${\bf M}=M_{x}{\bf e}_{x}$, i.e. ${\bf H}=H_{x}{\bf e}_{x}$
and ${\bf B}=B_{x}{\bf e}_{x}$. By the standard minimization procedure of
the Gibbs free-energy ${\cal G}\left[ \Delta _{n},{\bf L,M},{\bf B};{\bf H}%
\right] $ with respect to $\Delta _{n},{\bf L,M},{\bf B}$, and by assuming
the continuum limit \cite{Lev}, one gets the complete set of equations for
these quantities as well as the Gibbs free-energy of the vortex ${\cal G}%
_{v} $ - see also \cite{BuBuKuPaAdvances}, \cite{Buzdin}. 
\begin{equation}
(\delta +4\pi )M_{x}-\gamma L_{z}-B_{x}=0.  \label{mlb}
\end{equation}
The Maxwell equation for the magnetic field ${\bf B}$ reads 
\begin{equation}
\mathop{\rm rot}%
({\bf B}-4\pi {\bf M})=\frac{4\pi }{c}{\bf j}_{s},  \label{rotB}
\end{equation}
where 
\begin{equation}
{\bf j}_{s}={\bf j}_{ab}+{\bf j}_{z}.  \label{js}
\end{equation}
The in-plane current ${\bf j}_{ab}$ is given by Eq.(\ref{jab}) while ${\bf j}%
_{z}$ is the Josephson current between planes 
\begin{equation}
{\bf j}_{z}=j_{c}{\bf e}_{z}\sin \chi _{n,n+1}.  \label{jzet}
\end{equation}
The phase $\chi _{n,n+1}$ is given by Eq.(\ref{gphase}). In the following we
assume that the vortex axis, ${\bf B}$ and the external field ${\bf H}$ are
along the $x$-axis. By the standard procedure we get the equation for the
single vortex (centered on the origin) 
\begin{equation}
\lambda _{ab}^{2}\frac{\partial ^{2}B_{x}}{\partial z^{2}}+\lambda _{J}^{2}%
\frac{\partial ^{2}B_{x}}{\partial y^{2}}-\frac{B_{x}}{p}=0.  \label{Bvortex}
\end{equation}
The parameter $p=\delta /(\delta +4\pi )$ takes into account the additional
screening due to the appearance of the WF order. After straightforward
transformations the Gibbs energy of the vortex, ${\cal G}_{v}$, has the form 
\begin{equation}
{\cal G}_{v}=\frac{p^{2}}{8\pi }\int dxdy\left\{ B_{x}^{2}+\lambda
_{ab}^{2}\left( \frac{\partial B_{x}}{\partial z}\right) ^{2}+\lambda
_{J}^{2}\left( \frac{\partial B_{x}}{\partial y}\right) ^{2}\right\} -\frac{%
\Phi _{0}\widetilde{H}_{c1}}{4\pi },  \label{Gv}
\end{equation}
where 
\[
\widetilde{H}_{c1}=H_{ext}+4\pi pM_{\eta }^{0}. 
\]
$M_{\eta }^{0}$ is approximately given by Eq.(\ref{wf}).

The Eq.(\ref{Bvortex}) can be generalized by taking into account 
the nonlinear core effects \cite{Lev}. In that case the solution is 
\begin{equation}
B_{x}(y,z)=\frac{\Phi _{0}}{2\pi p\lambda _{ab}\lambda _{J}}K_{0}(\frac{R}{p}%
).  \label{BLond}
\end{equation}
where $R=\sqrt{y^{2}/\lambda _{J}^{2}+z^{2}/\lambda _{ab}^{2}}$ and $K_{0}$
is the Bessel function of the zeroth order of an imaginary argument. Inserting
such a solution into Eq.(\ref{Gv}) a straightforward calculation gives the
lower critical field $\widetilde{H}_{c1}$ from the condition ${\cal G}_{v}=0$
\begin{equation}
H_{ext}+4\pi pM_{x}^{0}=\widetilde{H}_{c1}\approx \frac{p\Phi _{0}}{4\pi
\lambda _{J}\lambda _{ab}}\ln \frac{\lambda _{ab}\sqrt{p}}{d},  \label{hc1}
\end{equation}
where $M_{x}^{0}=(\gamma /\delta )L_{z}$. We stress that the logarithmic
factor in Eq.(\ref{hc1}) is due to the nonlinear core effects of the
Josephson vortex \cite{Lev}. Note that in systems with $T_{N}\gg \theta _{em%
\text{ }}$, like in RuSr$_{2}$GdCu$_{2}$O$_{8}$, one has $p\sim 1$.

From (\ref{hc1}) it is seen that for 
\begin{equation}
M_{x}^{0}>\frac{\Phi _{0}}{16\pi ^{2}\lambda _{J}\lambda _{ab}}\ln \frac{%
\lambda _{ab}\sqrt{p}}{d}  \label{spvst}
\end{equation}
spontaneous vortices appear in the system. This condition is more easily
realized near the ``$0$''-to-``$\pi $'' transition (decoupling) line $%
T_{d}(h)$, i.e. when $\lambda _{J}$ is significantly increased. This means
that in systems where the exchange parameter fulfills the condition $%
0.87T_{c0}<h<3.77T_{c0}$ for s-wave pairing, while for d-wave pairing $%
0.87T_{c0}$ is replaced by $0.6T_{c0}$, then by lowering the temperature the 
$\widetilde{H}_{c1}(T,h)$ shows pronounced nonmonotonic behavior reaching
minimum at the ``$0$''-``$\pi $'' boundary line as it is seen in Fig.3.
\newline
\epsfysize=3.1in 
\hspace*{0.5cm} 
\vspace*{0cm} 
\epsffile{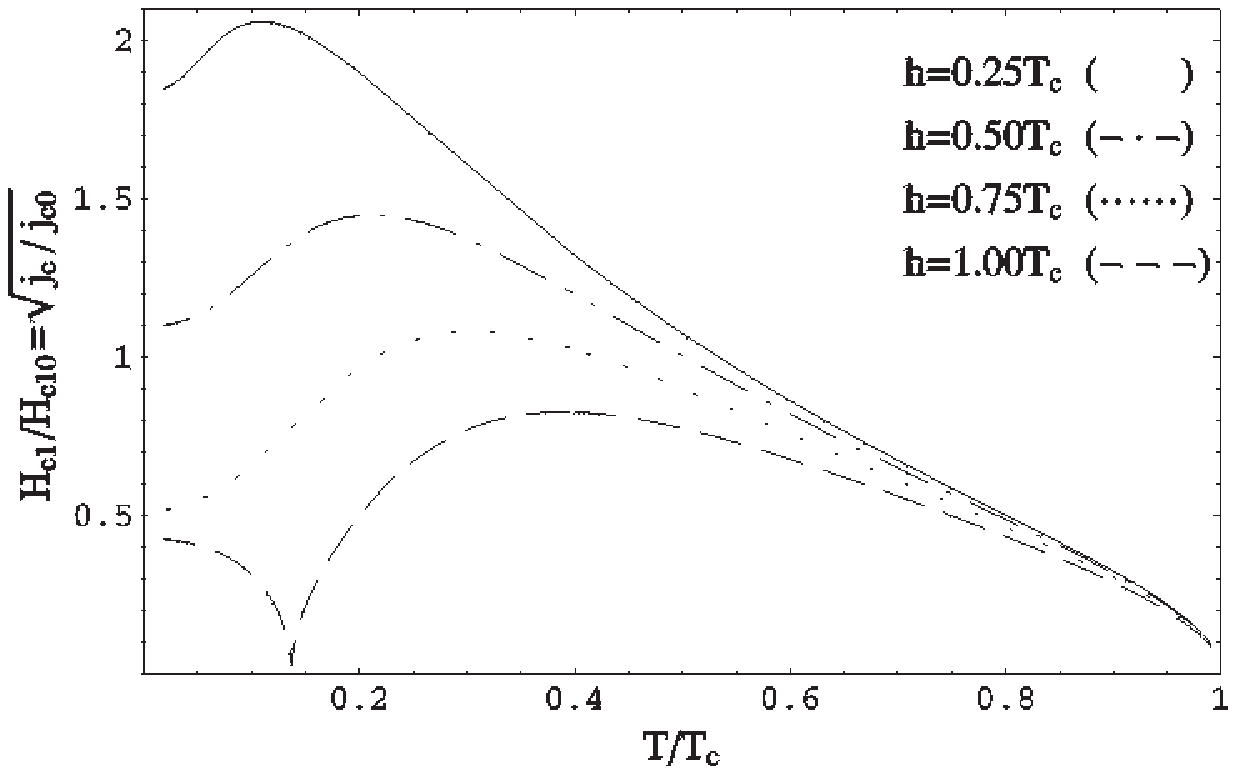} 
\newline
\mycap{{\bf FIGURE~3.} The $T$-dependence of the lower critical field, $\widetilde{H}_{c1}(T,h)$,
in the $ab$-plane for various $h$.We defined $\widetilde{H}_{c10}=\left(
p\Phi _{0}/4\pi \lambda _{J0}\lambda _{ab}\right) \ln \left( \lambda _{ab}%
\sqrt{p}/d\right).$ }

\section{Comparison with the experiment and discussion}

Let us discuss some relevant points related to the interpretation of the
obtained results on RuSr$_{2}$GdCu$_{2}$O$_{8}$.

(i) In order to analyze the magnetic properties the value of the exchange
field parameter $h=J^{ab}S_{%
\mathop{\rm eff}%
}$ is needed. If one takes the experimental value $S_{%
\mathop{\rm eff}%
}\sim 0.1$ one gets $\gamma /\delta \sim 0.1$ since $S_{%
\mathop{\rm eff}%
}\approx \gamma /\delta $. However, at present we do not know the relation
between $J^{ab}$ and $T_{N}$. As $d_{\text{Ru-Ru}}^{ab}\ll d_{\text{Ru-Ru}%
}^{c}$ one expects that the coupling of spins along the $c$-direction, $J^{c}
$ is much smaller than along the $ab$-plane, $J^{ab}$. In such a situation
one has $T_{N}\sim J^{ab}/\ln (J^{ab}/J^{c})$. Then, $J^{ab}>T_{N}$ and $%
h\sim 10-20%
\mathop{\rm K}%
$. We pay attention that there are evidences that in the underdoped HTS
materials d-wave pairing is realized \cite{Kulic}. In that case the point on
the phase diagram $j_{c}(T=0,h_{c})=0$ is realized for $h_{c}^{(d)}=0.6T_{c}$%
. According to the specific heat measurements \cite{Bernhard} in RuSr$_{2}$%
GdCu$_{2}$O$_{8}$ with $T_{c}=30-40%
\mathop{\rm K}%
$ this compound behaves like an underdoped HTS material. If it is so it
gives $h_{c}^{(d)}\sim 20%
\mathop{\rm K}%
$, i.e. $h$ is near to $h_{c}$ and a nonmonotonic behavior of $\widetilde{H}%
_{c1}$ is expected as shown in Fig.3.

(ii) If $M_{x}^{0}$ fulfills Eq.(\ref{spvst}) then there is a spontaneous
vortex state and the Meissner effect is absent. In opposite case the
Meissner state is realized.

(iii) It may happen that $M_{x}^{0}<\widetilde{H}_{c1}$ in some temperature
intervals and $M_{x}^{0}>\widetilde{H}_{c1}$ in the interval between, which
case corresponds to a reentrant behavior.

(iv) At present the origin and the magnitude of the parameter $\gamma $ in
Eqs.(\ref{mfree},\ref{wf}) is unknown. However, it may also happen that in
polycrystalline samples strains induce additional changes of this quantity.
A drastic case might be realized if the symmetry of the crystal implies that 
$\gamma =0$. Even in that case strains in samples, for instance the
component $\sigma _{xy}$, can induce a magnetic moment in piezomagnetic
systems, i.e. $M_{x}^{0}\sim \sigma _{xy}L_{z}$ thus producing weak
ferromagnetism. If strains in a sample are such that $M_{x}^{0}>\widetilde{H}%
_{c1}$ then the Meissner phase is not realized as reported in \cite{Chu}, 
\cite{Felner}. In such a way one could reconcile the opposite claims on
existence \cite{Braun}, \cite{Bernhard2} and nonexistence \cite{Chu}, \cite
{Felner}, of the Meissner phase in differently prepared samples of RuSr$_{2}$%
GdCu$_{2}$O$_{8}$.

(v) Based on the above analysis one expects that dynamical properties of
such a system are very exotic. For systems near the decoupling line $%
j_{c}(T_{d},h)=0$ there is a significant reduction of the Josephson plasma
frequency $\omega _{0,SF}\sim \sqrt{j_{c}}\ll \omega _{0,JJ}$ \cite{Doniach}
(compared to standard Josephson junction with $\omega _{0,JJ}$) for the
waves propagating along the $xy$-planes in the $S/F$ superlattice 
\[
\omega _{SF}^{2}=\omega _{0,SF}^{2}+v_{SF}^{2}q^{2}. 
\]
Due to the microscopic character of the S/F superlattice one expects that $%
v_{SF}^{2}\gg v_{JJ}^{2}$ where $v_{JJ}$ is the phase velocity for the
Josephson junction made from bulk superconductors. This means that in a S/F
superlattice, like for instance in RuSr$_{2}$GdCu$_{2}$O$_{8}$, it is
possible to tune $\omega _{0,SF}^{2}$ nonmonotonically and also to extract
the radiation with much higher intensity than in the single Josephson
junction. This is a matter of future investigations.

In conclusion, we have shown that in a S/F superlattice with the exchange
field $h\sim T_{c}$ acting in F-planes only a nontrivial and nonmonotonic
behavior of magnetic properties, like the lower critical field $H_{c1}$, is
realized. This property is due to the decrease of the effective Josephson
coupling between S-planes by increasing $h$.

{\it Acknowledgments} - One of us (M. L. K.) thanks C. Bernhard, H. F. Braun
and B. Keimer for useful discussion of their experimental results. M. L. K.
acknowledges the support of the Deutsche Forschungsgemeinschaft through the
Forschergruppe ``Transportph\"{a}nomene in Supraleitern und Suprafluiden''.
This work was also supported by the ESF ``Vortex'' Program.

\end{document}